\documentstyle[epsfig,12pt]{article}
\begin{document} 
\vspace*{-1in} 
\renewcommand{\thefootnote}{\fnsymbol{footnote}} 
\begin{flushright} 
DTP-98-30\\ 
TIFR/TH/98-18\\
June 1998 
\end{flushright} 
\vskip 65pt 
\begin{center} 
{\Large \bf $J/\psi$ production at the Tevatron and HERA:\\ 
the effect of $k_T$ smearing} \\ 
\vspace{8mm} 
{\bf 
K.~Sridhar${}^1$\footnote{sridhar@theory.tifr.res.in},
A.D.~Martin${}^2$\footnote{A.D.Martin@durham.ac.uk},   
W.J.~Stirling${}^{2,3}$\footnote{W.J.Stirling@durham.ac.uk} 
}\\ 
\vspace{10pt} 
{\sf 1) Department of Theoretical Physics, Tata Institute of 
Fundamental Research,\\  
Homi Bhabha Road, Bombay 400 005, India. 

2) Department of Physics, University of Durham, 
Durham DH1 3LE, U.K.\\ 
 
3) Department of Mathematical Sciences, University of Durham, 
Durham DH1 3LE, U.K.} 
 
\vspace{80pt} 
{\bf ABSTRACT} 
\end{center} 
\vskip12pt 
We study the effects of intrinsic transverse momentum smearing  
on $J/\psi$ production both at the Tevatron and at HERA. 
For the case of large-$p_T$ 
$J/\psi$ production at the Tevatron, the effects due to $k_T$ smearing are 
mild. On the other hand, inelastic $J/\psi$ photoproduction at HERA is very 
sensitive to the $k_T$ smearing and, in fact, with a reasonable value 
of $\langle k_T \rangle$ it is possible to resolve the large-$z$ 
discrepancy seen by comparing non-relativistic QCD (NRQCD) 
predictions with the HERA data. We 
conclude that, with the present kinematic cuts, photoproduction at 
HERA is not a good test of NRQCD. 
 
\setcounter{footnote}{0} 
\renewcommand{\thefootnote}{\arabic{footnote}} 
 
\vfill 
\clearpage 
\setcounter{page}{1} 
\pagestyle{plain} 
Over the last few years, there has been a considerable advance in the 
understanding of quarkonium physics due to the development of the 
non-relativistic effective field theory of QCD, called non-relativistic 
QCD (NRQCD) \cite{bbl}. The Lagrangian for this effective theory is obtained 
from the full QCD Lagrangian by neglecting all states with momenta larger than 
a cutoff of the order of the heavy quark mass, $m$, and  
accounting for this exclusion by introducing new interactions in the  
effective Lagrangian, which are local since the excluded states 
are relativistic. Beyond the leading order in $1/m$ the effective theory is 
non-renormalisable. The scale $m$ is an ultraviolet cut-off for the 
physics of the bound state; however the latter is more intimately tied to the 
scales $mv$ and $mv^2$, where $v$ is the relative velocity of the quarks 
in the bound state. The physical quarkonium state  
admits of a Fock expansion in $v$, and 
it turns out that the $Q \bar Q$ states appear in either 
colour-singlet or colour-octet configurations in this series. 
Of course the physical state must be a colour-singlet, so that a  
colour-octet $Q \bar Q$ state is connected to the physical 
state by the emission of one or more soft gluons. In spite of 
the non-perturbative nature of the soft gluon emissions, 
the effective theory still gives  
useful information about the intermediate octet states. This is  
because the dominant transitions that occur from colour-octet to physical 
colour-singlet states are $via$ E$1$ or M$1$ transitions with 
higher multipoles being suppressed by powers of $v$. It then becomes 
possible to use the usual selection rules for these radiative transitions 
to keep account of the quantum numbers of the octet states, so that 
the production of a $Q \bar Q$ pair in a octet state 
can be calculated and its transition to a physical singlet state 
can be specified by a non-perturbative matrix element. The cross-section 
for the production of a meson $H$ then takes on the following factorised form: 
\begin{equation} 
   \sigma(H)\;=\;\sum_{n=\{\alpha,S,L,J\}} {F_n\over m^{d_n-4}} 
       \langle{\cal O}^H_\alpha({}^{2S+1}L_J)\rangle , 
\label{e1} 
\end{equation} 
where the $F_n$'s are the short-distance coefficients and the 
${\cal O}_n$ are local 
4-fermion operators, of naive dimension $d_n$, describing the long-distance 
physics. The short-distance coefficients are associated with the production 
of a $Q \bar Q$ pair with the colour and angular momentum quantum numbers 
indexed by $n$. These involve momenta of the order of $m$ or larger and 
can be calculated in a perturbation expansion in the QCD 
coupling $\alpha_s(m)$. 
The $Q \bar Q$ pair so produced has a separation of the 
order of $1/m$ which is pointlike on the scale of the quarkonium 
wavefunction, which is of order $1/(mv)$. The non-perturbative 
long-distance factor $\langle {\cal O}^H_n\rangle$ is proportional 
to the probability for a pointlike $Q \bar Q$ pair in the state $n$ 
to form a bound state $H$.  
 
\vskip10pt 
The existence of the colour-octet components of the quarkonium wave function  
is the new feature of the NRQCD approach. Before the development of NRQCD, 
the production and decay of quarkonia were treated within the framework 
of the colour-singlet model \cite{br, berjon}. In this model, it is assumed 
that the $Q \bar Q$ pair is formed in the short-distance process in a 
colour-singlet state. The corrections from terms higher order in $v$ 
were neglected. While this model gave a reasonable description of low-energy 
$J/\psi$ data, it was known that it was incomplete because of an 
inconsistency in the treatment of the $P$-state quarkonia. This 
was due to a non-factorising infra-red divergence, noted first in 
the application of the colour-singlet model to $\chi_c$ decays \cite{bgr}, 
and the proper resolution of this problem was obtained only 
by including the colour-octet components in the treatment of the 
$P$-states \cite{bbl2}.  
The colour-octet components, however, had a more dramatic impact \cite{jpsi}  
on the phenomenology of $P$-state charmonium production at large $p_T$ at  
the Tevatron $p \bar p$ collider \cite{cdf} where the colour-singlet model 
was seen to fail miserably. 
While the inclusion of the colour-octet components for the $P$-states 
was necessary from the requirement of theoretical consistency, there 
was no such problem with the $S$ states because the corresponding amplitude 
was finite and the colour-octet components were suppressed 
compared to the colour-singlet component by $O(v^4)$. But the data 
on direct $J/\psi$ and $\psi'$ production at the Tevatron \cite{cdf} seem to 
indicate an important contribution from the colour-octet components 
for the $S$-states as well \cite{brfl}.  
 
\vskip10pt 
While it is clear that the correct description of the Tevatron large-$p_T$ 
data requires that the colour-octet components of the quarkonium 
wave function have to be taken into account, the major problem is that 
the corresponding long-distance matrix elements are {\it a priori} unknown and 
can be obtained only by fitting to the Tevatron 
data \cite{cho}. Clearly it is important to have other tests of NRQCD,  
so that the matrix elements which are obtained from the Tevatron data can be 
determined independently in other experiments.  
 
\vskip10pt 
One important cross-check is the inelastic photoproduction of $J/\psi$ at 
the HERA $ep$ collider \cite{hera}. The inelasticity of the events is 
ensured by choosing 
$z \equiv p_p \cdot p_{J/\psi} /p_p \cdot p_{\gamma}$ to be sufficiently 
smaller than one. An upper limit of $z \sim 0.8$ would seem to be an 
appropriate choice \cite{mns}. An additional cut needs to be imposed on 
the $p_T$ of the $J/\psi$ to ensure that the production process 
occurs at short distance. In the HERA experiments, the $z$ distributions 
have been studied using $p_T > 1$~GeV. The surprising feature of the  
comparisons \cite{ck, kls} of the NRQCD results with the data from HERA  
is that the colour-singlet model prediction is in agreement with the 
data while including the colour-octet component leads to violent 
disagreement with the data at large $z$. 
While the colour-singlet cross section dominates in most of the 
low-$z$ region, the colour-octet contribution increases steeply in the  
large-$z$ ($0.8<z<0.9$) region and this rise is not seen in the data.  
In these comparisons, the values of the non-perturbative matrix elements 
are taken to be those determined from a fit to the Tevatron large-$p_T$ 
data. Naively, one would think that this points to a failure of NRQCD. But 
this conclusion is premature and, in our opinion, incorrect. The reason 
is that while at the Tevatron the measured $p_T$ of the $J/\psi$ is 
greater than about 5~GeV, at HERA the $p_T$ can be as small as 
${\cal O}(1)$~GeV.  At such small 
values of $p_T$ (and also for $z$ very close to unity), there could be 
significant perturbative and non-perturbative soft physics effects. 
One way to parametrise these effects is to include the effects of the 
transverse momentum smearing of the partons inside the proton. This 
is the purpose of the present paper. We study the effects of the 
parton transverse momentum, $k_T$, on the $J/\psi$ distributions both at 
the Tevatron and at HERA. 
We demonstrate that the $z$ distribution measured at HERA is  
particularly sensitive to the effects of $k_T$ smearing, and argue that 
inelastic photoproduction at HERA, with the present kinematic cuts, 
is not a clean test of NRQCD. Other effects such as soft-gluon resummation 
\cite{softg} and the breakdown of NRQCD factorisation near $z=1$ \cite{wise} 
have been discussed in the context of this discrepancy. 
 
\vskip10pt 
In our work, we first consider the effects of $k_T$ smearing 
on the extraction of the non-perturbative matrix elements from 
the large-$p_T$ direct $J/\psi$ production data from the CDF experiment 
at the Tevatron. Before we discuss the fits, we recall the inputs that 
go into the theoretical computations of Ref.~\cite{cho} where 
these fits are performed in the absence of smearing. The direct 
$J/\psi$ production (i.e. $S$-state production, with the $P$-wave 
contributions removed) cross section in the NRQCD approach  
receives contributions from 1) the colour-singlet ${}^3S_1^{[1]}$ 
channel, 2) the colour-octet ${}^3P_J^{[8]}$ connected to the 
physical $J/\psi$ state by an E1 transition, 3) the colour-octet 
${}^1S_0^{[8]}$ channel which makes a M1 transition to the 
physical state, and 4) the colour-octet ${}^3S_1^{[8]}$ channel 
which connects to the $J/\psi$ state by a double E1 transition.  
All three colour-octet channels contribute at $O(v^7)$. The 
non-perturbative parameter for the colour-singlet channel (i.e. 
the radial wave-function at the origin) can be extracted from 
$J/\psi$ leptonic decay or estimated from potential model 
calculations. Given this input, the three non-perturbative 
parameters $\langle {\cal O}( {}^3P_J^{[8]}) \rangle$,  
$\langle {\cal O}( {}^1S_0^{[8]}) \rangle$,  
$\langle {\cal O}( {}^3S_1^{[8]}) \rangle$ (which we call 
matrix elements $M_1$, $M_2$ and $M_3$ respectively) have to be extracted 
by fitting to the CDF data. It turns out that for $p_T$ 
values greater than about 4 GeV, the $p_T$ dependence of 
the short-distance coefficients corresponding to the 
${}^3P_J^{[8]}$ and the ${}^1S_0^{[8]}$ channels are 
identical. The ${}^3S_1^{[8]}$ channel on the other  
hand has a different $p_T$ distribution: the reason 
for this is that a fragmentation-type diagram where 
a single gluon attaches itself to the $c \bar c$ pair 
is present only for this channel and not for the 
${}^3P_J^{[8]}$ and the ${}^1S_0^{[8]}$ channels.  
Due to the existence of this fragmentation-type 
diagram, the contribution from the ${}^3S_1^{[8]}$ channel  
dominates at large $p_T$ and so it is possible to use 
the experimental data points in this $p_T$ region to 
fit the value of the corresponding non-perturbative 
matrix element $M_3$. On the other hand, the contributions 
due to the ${}^3P_J^{[8]}$ and the ${}^1S_0^{[8]}$ channels are 
dominant at low $p_T$ and one can use data the points at the 
lower end of the measured $p_T$ spectrum to fit a linear combination 
of $M_1$ and $M_2$. The linear combination turns out to 
be $M_1/m_c^2 + M_2/3$. We use the phrase `fragmentation-type' 
guardedly. The subprocess cross sections to which we refer 
here are obtained from fixed-order perturbation theory 
calculations. In a genuine fragmentation calculation, 
the effects of Altarelli-Parisi evolution of the fragmentation 
functions are taken into account, and these go beyond  
fixed-order perturbation theory. In comparing the  
${}^3S_1^{[8]}$ prediction to the large-$p_T$ data, 
one must correct for the effect of this evolution,  
and the prescription for this correction (as given in 
Ref.~\cite{cho}) is to multiply the number obtained from the 
fixed order calculation with the correction factor $R$, 
where $R$ is the ratio of the cross sections (computed using 
the fragmentation approach) at the scales $p_T$ and $m_c$ 
i.e. with and without the $Q^2$ evolution of the fragmentation 
functions. 
 
\begin{figure}[t] 
\begin{center}
\mbox{\epsfig{figure=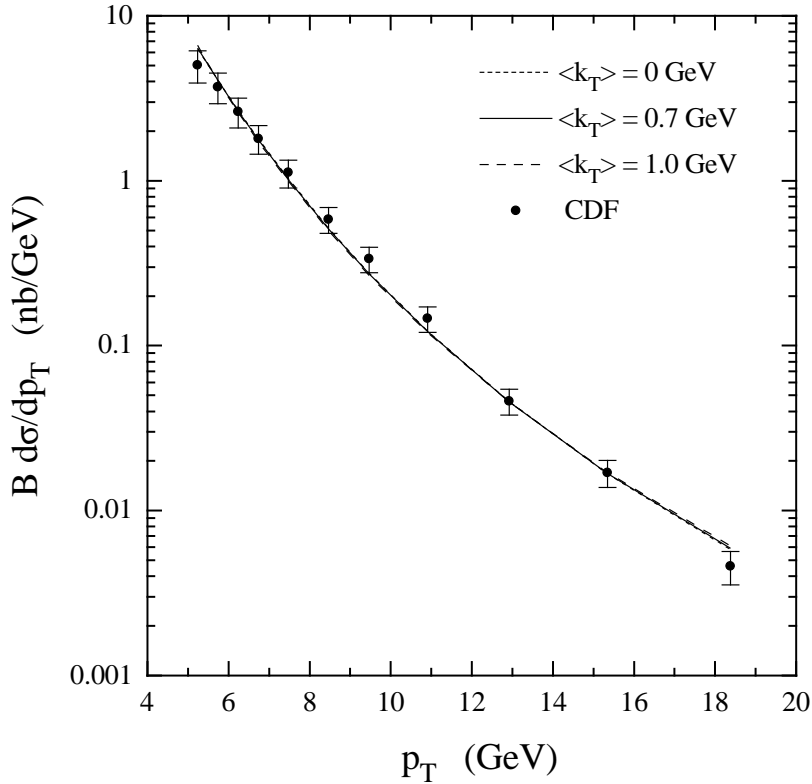,width=350pt}}
\caption{The CDF data \protect\cite{cdf} for
$Bd\sigma/dp_T$ (in nb/GeV) for $J/\psi$ production  
at 1.8~TeV with $-0.6 \le \eta \le 0.6$, compared to the model
predictions with  $\langle k_T 
\protect\rangle =0,\ 0.7,\ 1.0$~GeV, respectively.}
\end{center}
\end{figure} 
 
For the Tevatron, we perform similar fits to those described above 
except that we now include the effects of intrinsic transverse 
momentum smearing. The inclusion of $k_T$ smearing in large-$p_T$ 
reactions is done in the standard fashion, see for example Ref.~\cite{smear}. 
Each incoming parton is given an intrinsic transverse momentum,
with a gaussian distribution whose width is an adjustable parameter.
The gaussian shape is motivated by measurements of the transverse
momentum distribution of $l^+l^-$, $\gamma\gamma$ and $\pi\pi$
pairs produced in hadron-hadron collisions.
The cross section with the inclusion of smearing is therefore
given by
\begin{eqnarray} 
&&E{d^3\sigma \over dp^3}(p \bar p \rightarrow J/\psi X) 
=  \sum_{a,b,c=q,g}
 \int d^2k_{Ta} \int d^2k_{Tb}\int dx_1 \int dx_2 \nonumber \\
&& f_{a/p}(x_1, \vec{k}_{Ta}, Q^2) f_{b/\bar p}(x_2, \vec{k}_{Tb}, Q^2)  
 {1 \over \pi} {d\hat \sigma \over d \hat t}(ab \rightarrow {}^{2S+1}L_J c) 
\delta (\hat s + \hat t + \hat u - M^2)  , 
\label{e3} 
\end{eqnarray} 
where  
\begin{equation} 
f_{a/p}(x_1, \vec{k}_{Ta}, Q^2) = {F_{a/p}(x_1,Q^2)D(\vec{k}_{Ta}) \over
x_1^2 + (4 k_{Ta}^2 / s)}
\end{equation} 
and similarly for $ f_{b/\bar p}(x_2, \vec{k}_{Tb}, Q^2)$. $F_{a/p}$ is 
the usual parton distribution, and the function $D(\vec{k}_T)$ is the  
intrinsic transverse momentum distribution
\begin{equation}
 D(\vec{k}_T) = {b^2 \over \pi} {\rm exp}(-b^2 k_T^2) , 
\end{equation} 
with
\begin{equation} 
\langle k_T \rangle = {\sqrt{\pi} \over 2b} ,
\end{equation} 
and normalised such that
\begin{equation} 
\int d^2k_T D(\vec{k}_T) = 1. 
\end{equation} 
For the numerical phase-space integration, it is convenient to 
use the change of variables 
\begin{equation} 
k_{Ti} = \kappa \left({\rm ln} {1 \over q_i}\right)^2 ,\qquad (i=a,b)
\end{equation} 
so that
\begin{equation} 
\int_0^{\infty} k_{Ti} dk_{Ti} \rightarrow  \int_0^1  
dq_i {\kappa^2 \over 2q_i}  . 
\end{equation} 
In terms of the variable $q_i$ the Gaussian distribution becomes 
\begin{equation} 
D(q_i)= {b^2 \over \pi} \left( q_i \right)^{b^2\kappa^2} .
\end{equation} 
By choosing $\kappa$ such that $b \kappa $ is slightly larger 
than unity, we obtain a relatively smooth integrand.
 
\vskip10pt 
The results of the effects of including intrinsic $k_T$ for $J/\psi$ 
production at the Tevatron are shown in Fig.~1, where we consider the
$J/\psi$ $p_T$ distribution, $B d\sigma/dp_T$, with
$B$  the branching ratio of the $J/\psi$ into $\mu^+ \mu^-$.
The cuts on the pseudorapidity of the $J/\psi$ are those used
by the CDF experiment i.e. $-0.6 \le \eta \le 0.6$.  
In our computations we use MRS(D$_-^{\prime}$) parton
densities \cite{mrs} with factorisation
scale $Q=M_T\equiv \sqrt{M_\psi^2 + p_T^2}$.
We present the results for three
different values of $\langle k_T \rangle$, {\it viz.} $\langle k_T \rangle
= 0,\ 0.7,\ 1.0~{\rm GeV}$. The choice $\langle k_T \rangle = 0$
corresponds to no smearing.
For the colour-singlet parameter, $\langle {\cal O}( {}^3S_1^{[1]}) \rangle$,
we use the value 0.81 GeV$^3$. This corresponds, of course, to the case
where there is no smearing. For the case when smearing is included
we need to know the effect of the smearing on the value of this parameter.
For this purpose, we use the data on $J/\psi$ production
from the EMC collaboration \cite{emc} in the limited $z$ range  
$0.3<z<0.6$. In this range, the colour-singlet contribution  
gives almost the entire cross-section \cite{kls}. We can therefore
use these data to extract the effect of the smearing on the singlet
wave-function. For $\langle k_T \rangle = 0.7,\ 1.0$~GeV, we find
the values to be 0.98~GeV$^3$ and 1.41~GeV$^3$ respectively.
The values of the non-perturbative octet parameters extracted for the 
three choices of $\langle k_T \rangle$ are
tabulated in Table~1. We see that the effect of the
$k_T$ smearing on the parameters extracted from the large-$p_T$ CDF 
data is very modest. Fig.~1 shows that the fits to the data when $k_T$
smearing is included are very good and comparable in quality to the 
case $\langle k_T \rangle = 0$.  
  
\begin{table}[htbp] 
\begin{center} 
\begin{tabular}{|c|c|c|} 
\hline 
$\langle k_T \rangle$ (GeV)& $M_1/m_c^2+M_2/3$ (GeV)${}^3$& $M_3$
(GeV)${}^3$\\  
\hline 
0.0 & $(3.14 \pm 0.58) \times 10^{-2}$ & $(1.26 \pm 0.33)
\times 10^{-2}$\\ 
0.7 & $(2.82 \pm 0.47) \times 10^{-2}$ & $(1.35  \pm 0.30)
\times 10^{-2}$\\ 
1.0 & $(2.35 \pm 0.39) \times 10^{-2}$ & $(1.50 \pm 0.29)
\times 10^{-2}$\\ 
\hline 
\end{tabular} 
\caption{The values of the colour-octet non-perturbative matrix
elements determined from the CDF data on $J/\psi$ production,
for three choices of intrinsic initial state transverse momentum.}
\end{center} 
\end{table} 
  
Taking these fitted values of the parameters, we next consider
inelastic $J/\psi$ photoproduction at HERA. For photoproduction, the
formalism for the inclusion of smearing effects is similar 
to that used for the  $p \bar p$ collisions described
above, except that now we have only one hadron in the initial
state. We take the same choice of parton distributions, scales etc.
as used in the Tevatron fits.
We compute the $z$ distribution for a photon-proton
centre-of-mass energy $\sqrt{s_{\gamma p}} =100$~GeV,
imposing a transverse momentum cut $p_T > 1$~GeV.
Again we present results for $\langle k_T \rangle =0,\ 0.7,\ 1.0$~GeV.
For each choice of $\langle k_T \rangle$, the values of the octet
non-perturbative matrix elements are taken from Table~1, and
the singlet matrix elements are taken to be the same as that
used in the Tevatron fits.
\begin{figure}[t] 
\begin{center}
\mbox{\epsfig{figure=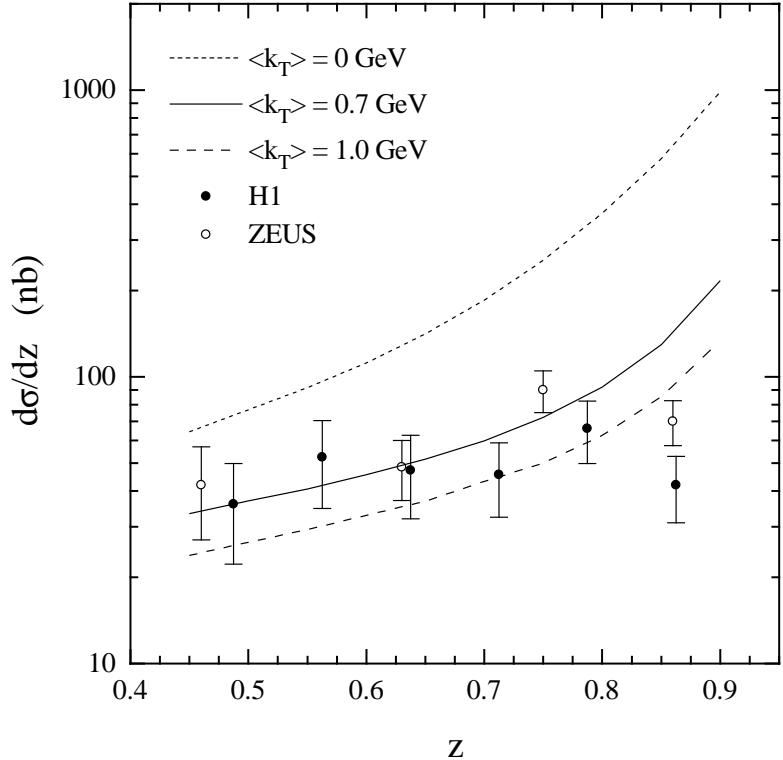,width=350pt}}
\caption{The HERA data \protect\cite{hera} for $d\sigma/dz$ (in nb)
for $J/\psi$ production
at $\protect\sqrt{s_{\protect\gamma p}}=100$~GeV with $p_T>1$~GeV,
compared with model predictions for three choices of the intrinsic
transverse momentum distribution, namely
$\protect\langle k_T \protect\rangle =0,\ 0.7,\ 1.0$~GeV.}
\end{center}
\end{figure} 
 
With these inputs, we compute the $z$ distribution for the
three different values of $\langle k_T \rangle$. The results are 
compared with HERA data \cite{hera} in Fig.~2.
In the absence of smearing, $\langle k_T \rangle =0$, we see that
 the colour-octet
component makes a large contribution at $z$ close to 1 which is not
supported by the data. However the introduction of $k_T$ makes a substantial
change to the octet contribution.
Whereas the effect of $k_T$-smearing is very small for large-$p_T$
production at the Tevatron, these effects are found to be
very important for $J/\psi$ production at HERA. In particular, smearing
signficantly reduces the size of the cross section (mainly due to the
smearing function itself, but also due to the smaller octet matrix elements,
see Table~1.) The $z$ distribution also becomes flatter,
in better agreement with the HERA data. In fact, excluding the highest
two data points at $z \approx 0.86$ we find that the $\chi^2$ is minimised
for $\langle k_T \rangle \sim 0.7$~GeV and the resulting description
of the data is very good. It is perhaps not surprising that the highest-$z$
data points are not exactly accounted for, since in this region
other effects such as contamination from elastic scattering or
breakdown of NRQCD factorisation may be important.
 
In conclusion, while a direct comparison of the predictions of NRQCD 
with the $z$-dependence of the inelastic photoproduction cross section
for $J/\psi$ at HERA show a marked disagreement between the two, we
argue that such a comparison is misleading. The inelastic photoproduction
process does not provide a clean test of NRQCD  because in order to
have a sizeable event rate, a very low $p_T$-cut ($\sim 1$~GeV) is  
necessary in the HERA experiments. At such low values of $p_T$ (and
for values of $z$ reasonably close to one), we find that the effect of 
$k_T$-smearing is important and that, indeed, for
$\langle k_T \rangle \sim 0.7$~GeV, the discrepancy between theory and
experiment is no longer observed. On the other hand, the inclusion 
of $k_T$ smearing has a very modest effect on the large-$p_T$ $J/\psi$ 
data from the Tevatron. Better tests of NRQCD may be obtained
by studying other observables at the Tevatron itself, such as the study
of the polarisation of the produced $J/\psi$ \cite{trans} or the
production of other charmonium resonances \cite{self,mps} whose
cross-section can be predicted in NRQCD. 

\vskip 0.5cm
 
\noindent{\bf Acknowledgements}

\noindent K.S. wishes to thank the Durham Centre for Particle
Theory for hospitality for his visit. He would also like
to thank U.K. PPARC for a Fellowship during his stay in Durham.
We thank V.~Papadimitriou for providing us with the experimental
numbers from CDF and Beate Naroska for  HERA data.

\end{document}